\documentclass[conference]{IEEEtran}

\ifCLASSINFOpdf
\else
\fi
 \usepackage[usenames,dvipsnames]{color} 
\usepackage{amssymb}
\usepackage{amsmath}
\usepackage{textcomp} 
\usepackage{xspace,textcomp}
\usepackage{rotating}
\usepackage{focus} 

\newcommand{\focust}{\textsc{Focus}$^{ST}$}

\begin{document}
\IEEEoverridecommandlockouts
\IEEEpubid{\makebox[\columnwidth]{~\copyright~2015 IEEE \hfill} \hspace{\columnsep}\makebox[\columnwidth]{ }}

\renewcommand{\uft}{\small\textsf{ft}}
\renewcommand{\urt}{\small\textsf{rt}}
\newcommand{\fft}{\footnotesize\textsf{ft}}
\newcommand{\frt}{\footnotesize\textsf{rt}} 
\newcommand{\fasm}{\small\textsf{asm}}
\newcommand{\fgar}{\small\textsf{gar}} 
\newcommand{\fntrue}{\footnotesize\textsf{true}}
\newcommand{\fnfalse}{\footnotesize\textsf{false}}
 
\renewcommand{\ulocal}{\small\textsf{local}\xspace}
\renewcommand{\uuniv}{\small\textsf{univ}\xspace}
\renewcommand{\uinit}{\small\textsf{init}\xspace}
\renewcommand{\uinitP}{\small{\textsf{initProcess}\xspace}} 
\renewcommand{\nentry}{\textsf{entry}~ }
\renewcommand{\nexit}{\textsf{exit}~~ } 

\newcommand{\sculocal}{\scriptsize\textsf{local}\xspace}
\newcommand{\scnlocal}{\tab{\sculocal}}
\newcommand{\scuuniv}{\scriptsize\textsf{univ}\xspace}
\newcommand{\scnuniv}{\tab{\scuuniv}}
\newcommand{\scuinit}{\scriptsize\textsf{init}\xspace}
\newcommand{\scuinitP}{\scriptsize{\textsf{initProcess}\xspace}}  
\newcommand{\scninitP}{\\ \zeddashline \tab{\scuinitP}\t1 ~}
\newcommand{\scninit}{\\ \tab{\scuinit}}
 
\title{Reconciling a component  and process view}
 
\author{\IEEEauthorblockN{Maria Spichkova  and
Heinz Schmidt }
\IEEEauthorblockA{RMIT University, Melbourne, Australia\\
 Email:  \{Maria.Spichkova,Heinz.Schmidt\}@rmit.edu.au } 
}

\maketitle
\pubid{\copyright~2015 IEEE}
 
\begin{abstract}%
In many cases we need  to represent on the same abstraction level not only system components  
but also processes within the system, and if for both representation different frameworks are used, 
the system model becomes hard to read and to understand. 
We suggest a solution how to cover this gap and to  reconcile   component and process  views on system representation:  
a formal framework  that gives the advantage
of solving design problems for large-scale component systems.
\end{abstract}

\section{Introduction}

Component-based software engineering is one of the largest fields of software and system engineering, 
however, in many cases we  need to represent on the same abstraction level not only system components  and the data flows between them 
but also processes within the system. 
Even if  the common practice to model parts of a system is to use the component view, 
the representation of  system behaviour by modelling  processes within the system 
becomes more and more important: nowadays the process view and 
the data flow representation are a typical part of the development of interactive or reactive systems. 
Having a process view, we can abstract from several aspects of the data flows by focusing on the control flow within the system, 
which gives us the advantage of comprehensible representation even in the case of  large-scale systems. 
However, if we need to have both, process and component, views on the system to get a comprehensive system model, the gap between these views can 
reduce to zero the benefit of having both kinds of representation.  
To cover this gap, we present a formal model of processes which is compatible with the component view:   
modelling both components and processes within  the same  framework, 
we not only increase  the readability of a system model but also can easier ensure consistency among these different views on a system. 

On the one hand, this concept  can also be related  to  the  IEC 61499 standard  developed 
as a technology for distributed automation systems with the decentralised and distributed control logic 
(cf. \cite{IEC614991,Vyatkin2011_IEC61499}). %
This standard is oriented on the development of  reusable modules for industrial control applications,
 and purposes to use function blocks as the basic constructs. 
 Each function block corresponds to an abstract representation of  a functional unit of software, 
 where local data and the behavioural specification  are encapsulated within an event-data interface.
In IEC 61499, modular components have an event signal interface (representing a control flow)  and data ports (representing a data flow),
  and are coupled in a hierarchical manner to arrange more complicated,  compositional blocks. 
 Thus, the suggested approach can be seen as a formalisation of  the main parts of the IEC 61499 standard, 
 however, due to suggested syntax, we can also switch from an event-based specification to a  time-triggered one -- in both cases we have a trigger of some kind, the difference solely is whether the trigger is an explicit data/control signal  or an information about the current time in a system. 
On the other hand, specifying a process view on a system, it is desirable to have a possibility of  a flexible translation from/to a common Petri Net notation (cf. eg.,~\cite{ReisigPetriNets,Salimifard}), which allows to focus on the control flow analysis within a system and is mainly recognised as a modelling language for process representation. 
Our approach enables schematic translation between the suggested process view representation and the Petri Net language.

%
\section{Related Work}

 Component-based software engineering utilises a well-defined
composition theory  to enable the prediction of such properties 
as performance and reliability.
This is one of the largest fields of software and system engineering. 
There are many approaches on component-based development covering different aspects and focusing on requirements, quality, timing properties etc.  
(cf. e.g., \cite{CBSQ2003,Broy2010MSS,cocome}). %
Several  component-based prediction approaches, 
e.g.\ Palladio~\cite{Kapova2010}, %
CB-SPE~\cite{CB-SPE}, 
ROBOCOP~\cite{Robocop}
 derive the benefits of  reusing well-documented component specifications  (cf. also a survey in~\cite{Becker06performanceprediction}). 
 In our approach we focus on the questions of combination of component/data flow  and process views, 
 to reuse most of the advantages of both representations and to avoid gaps in having these representations as unconsolidated ones.

There is a large variety of approaches on process representation  
which have a lot of similarities as well as differences  in many aspects such as (co-)algebraic view, 
 composition types, kinds of structuring, representation of time, separation of different kinds of flow, etc.   
An informal way to represent processes is used in the UML (Unified Modeling Language, \cite{UML_2005}): 
the concept of activity diagrams supports the specification of control flow in terms of choice, iteration, and concurrency. 
However, 
there is a number of approaches, e.g., \cite{Storrle04semanticsand}, which aim to  formalise the UML semantics in different ways. 
A co-algebraic view on process modelling gives, e.g., the  coordination language Reo~\cite{reo2012}  %
a channel-based modelling language
that introduces various types of channels and their composition rules. By composing Reo channels, we can
specify connectors to realise some behavioural protocols. The main concepts used in this language are the service synchronisation and
the data flow constraints.

\newpage
Many process description techniques are also based on the ideas of Petri Nets~\cite{ReisigPetriNets,Salimifard}. For example,  
YAWL (Yet Another Workflow Language, cf. \cite{Aalst03yawl_yet}) was developed by taking Petri Nets as a starting point and adding new  mechanisms
on  the workflow patterns. 
A number of architecture description languages (ADLs) have been
developed to specify compositional views of a system on an abstract
level, e.g., the TrustME ADL~\cite{Schmidt2001trustByContracts}, which
combines software architecture specification approaches with ideas of
design-by-contract.  This approach allows capturing of complex
behavioural interaction patterns, synchronous and asynchronous,
between large-scale components of software and systems architectures.

Other approaches  formalise work flows using the concept of process algebras~\cite{HandbookProcessAlgebra}.
The most famous of them are
Bergstra's Algebra of Communicating Processes~\cite{ACP}, 
Hoare's approach on Communicating Sequential Processes (CSP)~\cite{CSP}, 
Milner's Calculus of Communicating Systems (CCS, cf.~\cite{CCS}), and variants thereof. 
This kind of techniques do not provide the high level of abstraction which is very important 
in the early phases of system development. 
Nevertheless, general ideas of process algebras influence many other methods, and 
there are works aimed to solve the problem with abstract view, e.g., by adding graphical notations like in~\cite{hilderink2003}.

 
\section{Formal Model of Processes}
\label{sec:process}

 From the large collection of process description techniques, we  chose the process language described in \cite{leux2010}
as the most suitable for our purpose  to embed the process view into the component representation: 
this process language combines the concept of (de)activating processes via control points, firstly introduced in Petri Nets,  
with the idea of separation of data and control flow to enable a proper composition of processes. 
A process is understood there as ``an observable activity executed by one or several actors, which might be persons, components, technical systems, or combinations thereof'': it has one entry (activation, start) point and one exit (end) point, what also perfectly fits to the main ideas of the IEC 61499 standard: 
an \emph{entry point} is a special kind of input channel that activates the process (the functional block in IEC 61499), 
where an \emph{exit point} is a special kind of output  channel that is used to indicate that the process (computation in the fictional block) is finished. 
Our approach allows us to model elementary and composed processes in a formal way, 
 to argue about properties of composed systems, and easily switch from the  process view to a classical component view. 
 The hierarchical definition of a process gives many advantages for analysis, 
 the formal model of a process permits its formal verification, and, moreover, 
 provides a formal interpretation for the behaviour of a process as a special kind of a component.  
 
Formal specification frameworks  should include predefined templates and special alerts helping to avoid 
the omission of assumptions about the systemÕs environment. 
For this reason we specify every component  in terms of an assumption and a guarantee: 
whenever input from the environment behaves in accordance with the assumption, 
the specified component is required to fulfil the guarantee. 
Even the application of  specification templates can make the model development more understandable and 
more appropriate for safety-critical and large-scale component systems  \cite{Spichkova2013HFFM}.  
The main ideas presented in this paper are mostly language-independent, 
nevertheless we prefer to present them using an algebraic language 
\focust\cite{focusST}  inspired by \Focus\cite{focus}, 
a framework for formal specification and development of interactive systems. 
Another advantage is a well-developed theory of composition.

We specify for any process $P$ its entry and exit points by $Entry(P)$ and $Exit(P)$ respectively, and
represent a process $P$ (elementary or composed) by the corresponding component specification $PComp$, thus,  
$
  \semantics{P} = PComp  	
$.
For any process $P$ with syntactic interface $\nint{I_P}{O_P}$, where  
$I_P$ and $O_P$ are sets of input and output data streams respectively,  
we can specify   $I_{\semantics{P}} = \{ Entry(P) \} \cup I_P$ and $O_{\semantics{P}} = \{ Exit(P) \} \cup O_{P}$.

A process can be defined as an elementary or a composed one, where the composition of any two processes $P_1$ and $P_2$ 
can be sequential $P_1 ; P_2$, 
alternate $P_1 \oplus P_2$ or parallel $P_1 || P_2$, 
and for any process $P$ we can define repetitively composed process $P\circlearrowleft_{lpspec}$, 
where $lpspec$ denotes a loop specifier.  
We treat a process as a special kind of a component that has additionally two extra channels (one input and one output channel) 
which are used only to activate the process and to indicate its termination, i.e.\ to
 represents the entry and exit points of the process.

The formal correlation between the  definition of processes and components are presented below, 
separately for elementary and composite processes.  
\emph{Composite specifications} of processes (as well as of components)
are built hierarchically from elementary ones using 
constructors for composition, and can be represented in the 
\emph{graphical} or the \emph{textual} style. 

In this paper we use the following operators to present examples of process/component specifications:
{\small 
\begin{center}
\begin{tabular}{ c  l  }
~ $\nempty$ 		&  an empty stream ~
\\
~ $\angles{x}$		&  one element stream consisting of the element $x$ 
\\
~ $\nft{l}$ 		 	&   the first element of an untimed stream  $l$ ~
\\
~ $\ti{s}{i}$		  &  the $i$th time interval of the stream $s$ ~
\\
$\fmsg{n}{s}$ 		 &   
 $s$  can have at most $n$ messages at each time interval  ~
\\
\end{tabular}
\end{center}
}

\subsection{Model of an elementary process}
\label{sec:process_elem}

An elementary process corresponds to an elementary specification that has one special input channel \emph{start} of type \emph{Event} consisting of one element $\event$
 as well as one special output \emph{stop} of the same type 
 (input and output points of the process that corresponds to the signals  \emph{process is started} and \emph{process is  finished}).
Using the syntax proposed in this paper, we  specify the type of these  channels only implicitly, and need to have the following 
 extensions of a component to model a process: 
\begin{itemize}
\item 
Each input channel (except the activation signal channel) $c$ has a corresponding buffer (local variable) $cBuf$ of size one (one element buffer), which value will be taking into account, when starting the process.
\item
If the process is inactive, there are no values on its output channels.
\item
The component  gets a local variable $active$ of type $\Bool$ to represent whether the process is in \emph{active phase}.
\end{itemize}

\begin{figure}
 \centering%
\vspace{-5mm}
{\scriptsize
\begin{procF}{P \cbox{[start, stop]}}{Parameters}{timed}
\InOut{x_1: MI_1, \dots, x_n: MI_n}{y_1: MO_1, \dots, y_m: MO_m}\\
\scnlocal  x_{1}Buf \in MI_1; ...;  x_{n}Buf \in MI_n\\
\scninit  
    x_{1}Buf = BufInit_1; ...;  x_{n}Buf = BufInit_n\\
\scninitP InitValuesReqForEveryProcessRestart
\zeddashline
\uasm \t1 SomeAssumptions
\ngar 
 \cbox{1} 
~~
 PrEnding (\ti{x_1}{t}, \dots, \ti{x_n}{t}, x_{1}Buf, \dots,  x_{n}Buf)  ~\to~  
 \ti{ext}{t} = \angles{\event}  ~ \wedge\\ 
  PrCalcF(\ti{x_1}{t}, \dots, \ti{x_n}{t}, x_{1}Buf, \dots, x_{n}Buf, \\
   \t4   x_{1}Buf', \dots, x_{n}Buf',\ti{y_1}{t}, \dots, \ti{y_m}{t})  
 \\
 ~   \\
 \cbox{2} 
 ~~ 
 \neg PrEnding(\ti{x_1}{t}, \dots, \ti{x_n}{t}, x_{1}Buf, \dots,  x_{n}Buf)  ~\to~ 
 \ti{ext}{t} = \nempty  ~ \wedge\\
  PrCalc(\ti{x_1}{t}, \dots, \ti{x_n}{t}, x_{1}Buf, \dots, x_{n}Buf, \\
    \t4  x_{1}Buf', \dots, x_{n}Buf',\ti{y_1}{t}, \dots, \ti{y_m}{t})  \\
\end{procF}
}\vspace{-7mm}
    \caption{Specification of a process $P$}
    \label{fig:process}
\end{figure}

\noindent
We suggest the following framework for process specification. 
Assume a process $P$ has $n$ input channels $x_1, \dots, x_n$ and $m$ output channels $y_1, \dots, y_m$ 
(cf.  Figure \ref{fig:process} for a general specification and Figure \ref{fig:comp} for the corresponding component specification). 
Data types of input and output streams are denoted by $MI_1$, \dots, $MI_n$ and $MO_1$, \dots, $MO_m$ respectively. 
In the $\ulocal$-section of the specification  we introduce all the local variables used by the process as well as 
the buffer variables used to store the values of the latest inputs while the process is inactive.  
The initial values of buffers for the input channels $x_1, \dots, x_n$ are denoted by $BufInit_1$, \dots,  $BufInit_n$.  
A process can also have  a number of parameters which can be listed in parenthesis.

The specification section $\uinitP$ differs from the section $\uinit$ in the following sense:
everything that is defined within the $\uinit$ section must be initialised only once, in the beginning, 
where everything that is defined within the $\uinitP$ section must be initialised every time the process is (re)started, 
i.e. every time the value of the local variable $active$ is triggered from $\fnfalse$ to $\fntrue$ 
(in a process specification this trigger is used implicitly, where in a component specification we specify these changes directly).

To increase readability, we label all transitions in the state transition diagram:
dealing with specifications or real systems, where a diagram could be hardly readable due to its size 
and a large number of state transitions, we need to use another representation style. 
Each table line (in the case of a diagram, each transition) can be specified as a single formula in the \ugar-part of the specification, the rewriting scheme is straightforward.
In addition, we distinguish two types of the transition labels  by coloured representation: 
inputs and constraints on the current local variables' values  are marked blue,   
outputs and changes of  local variables' values %
are marked green.

\begin{figure}
 \centering%
{\scriptsize
\begin{specF}{PComp}{Parameters}{timed}
\InOut{start:Event; x_1: MI_1, ..., x_n: MI_n}
{stop: Event; y_1: MO_1, ..., y_m: MO_m}\\
\scnlocal active: \Bool; x_{1}Buf \in MI_1; ...;  x_{n}Buf \in MI_n\\
\scninit active = \nfalse;~
        x_{1}Buf = BufInit_1; ...  x_{n}Buf = BufInit_n
\zeddashline
\uasm  \t1 SomeAssumptions
\ngar 
 \lbox{1} 
~~ active = \ntrue ~\wedge~\\
   PrEnding(\ti{x_1}{t}, ..., \ti{x_1}{t}, x_{1}Buf, ...,  x_{n}Buf)  ~\to~ 
    \ti{ext}{t} = \angles{\event}  ~ \wedge\\
  PCalcF(\ti{x_1}{t}, ..., \ti{x_1}{t}, x_{1}Buf, ..., x_{n}Buf,  x_{1}Buf', ..., x_{n}Buf',\ti{y_1}{t}, ..., \ti{y_m}{t}) \\
  \wedge~  active' = \nfalse  \\
    ~\\
 \lbox{2} 
~~ active = \ntrue ~\wedge~\\
  \neg PrEnding(\ti{x_1}{t}, ..., \ti{x_1}{t}, x_{1}Buf, ...,  x_{n}Buf)  ~\to~ \ti{ext}{t} = \nempty  ~ \wedge\\
  PCalc(\ti{x_1}{t}, ..., \ti{x_1}{t}, x_{1}Buf, ..., x_{n}Buf,  x_{1}Buf', ..., x_{n}Buf',\ti{y_1}{t}, ..., \ti{y_m}{t})  \\
\wedge~ active' = \ntrue  
\\
   ~\\
 \lbox{3} 
~~
active = \nfalse ~\wedge~  \ti{ent}{t} = \nempty ~\to~\\
\ti{ext}{t} = \nempty  ~\wedge~ active' = active ~\wedge~ \ti{y_1}{t} = \nempty ~\wedge~ ... ~\wedge~ \ti{y_m}{t} = \nempty \\
~\\
\lbox{4} 
~~
  active = \nfalse ~\wedge~  \ti{ent}{t} \neq \nempty 
  ~\to~\\
    InitValuesReqForEveryProcessRestart~\wedge~ \\
    \ti{ext}{t} = \nempty  ~\wedge~ active' = \ntrue ~\wedge~ \ti{y_1}{t} = \nempty ~\wedge~ ... ~\wedge~ \ti{y_m}{t} = \nempty\\
~\\
 \lbox{5} 
~~
 active = \nfalse ~\wedge~ \ti{x_1}{t} \neq \nempty  ~\to~ 
 x_{1}Buf' = \nft{\ti{x_1}{t}} 
    \\
        \lbox{6} 
~~
 active = \nfalse ~\wedge~ \ti{x_1}{t} = \nempty ~\to~
  x_{1}Buf' = x_{1}Buf 
  \\
    ...
    \\
\lbox{5+2n} 
~
 active = \nfalse ~\wedge~ \ti{x_n}{t} \neq \nempty  ~\to~ 
 x_{n}Buf' = \nft{\ti{x_n}{t}} 
  \\ 
  \lbox{6+2n} 
~
   active = \nfalse ~\wedge~ \ti{x_1}{t} = \nempty ~\to~
  x_{n}Buf' = x_{n}Buf 
~\\
\end{specF}
}
\vspace{-7mm}
    \caption{Specification of a component, representing the process $P$}
    \label{fig:comp}
\end{figure}

The $\fasm$-part of the specification must contain all the assumption about the environment, i.e. 
all the properties of input streams which are necessary for the correct system behaviour. 
The $\fgar$-part of the specification contains the description of system behaviour:
the behaviour of any process in its \emph{active phase}.  
The condition of the process finishing is defined by the relation \emph{PrEnding} over the received input values.

The relation $PrCalcF$ describes the calculations  of the output and buffer values for the case $PrEnding$ holds,
 however, sometimes we can use the same predicate for both cases. 
By the relation $PrCalc$  we represent here all the calculations of the output and local values for the current step/time unit 
and of the buffer values for the next step -- they have to be performed during the time  process is active. 
In some cases we need to extend this predicate by calculations of some other local variables of the process.

Below we have presented a general specification of a process $P$ following by 
the corresponding component specification \emph{PComp}. 
The $1$st and $2$nd formulas in the component specification are almost equal to the formulas in the process specification: 
the constraints on the variable $active$ are now added explicitly.  
 The behaviour of any process in its \emph{inactive phase} is defined in the component specification by the formulas $3, \dots, 6+2n$. 
It is the same for any process, and is therefore omitted in process specifications.  
The only exception is the formula~$4$: initialisation of the values of local variables: if it is required for every restart of a process, 
the corresponding constraints should be moved to the $\uinitP$ section.

It is easy to see that a process and a classical component specification have a very similar structure and syntax, 
and one can easily change from one view to the other without any effort  and learning a new language. 
The same also holds for composed processes and components.

We suggest to represent  $PrCalc$  by a state transition diagram or the corresponding state transition table (also combining it with the representation of $PrCalcF$), 
because a graphical specification is, in general,  more readable than a plain text one. 
Here is applicable the idea of mode automata, which have a long history motivated by real-time design
practices and methods used in industry in connection with statecharts.
Maraninchi et al.~\cite{maraninchi03mode} capture %
the notion of modes formally for a practical extension of the
real-time synchronous language Lustre and include elements of the
well-known I/O-automata.  Mode automata define synchronous mode
automata as a hybrid between data-flow and transition systems, and in our case we need only a part of their approach: 
a process in our framework has only two modes, \emph{Active} and \emph{Inactive},  that correspond two possible values of the variable \emph{active}.  
To argue about a mode of a process $P$ at time interval $t$ 
we use the predicate $\nactive{P}{t}$ introduced below.  

In the stream representation we say that streams $x_{1}, \dots, x_{n}$ are \emph{disjoint}  
  (denoted by $\ndisjoint{x_{1}, \dots, x_{n}}$) iff 
on every time interval $i$ only one of these streams contains messages, i.e. 
$$
\forall t \in \Nat, i \in [1..n] : \ti{x_{i}}{t} \neq \nempty \to \forall j \neq i,\  j \in [1..n]:   \ti{x_{j}}{t} = \nempty
$$
We can extend this idea to the operation over components and processes (as a special kind of components). 
A component $C$ is \emph{active} on output stream $x \in \outstreams(C)$ on the time interval $t$ 
if on this time interval the stream $x$ is nonempty
$$
\nactiveX{C}{t}{x} \ndef \exists x \in \outstreams(C): \ti{x}{t} \neq \nempty
$$
and 
it is \emph{active only  on output stream x}  on the time interval $t$ 
if on this time interval all other its output streams are empty
$$
\nactiveXR{C}{t}{x} \ndef \exists!\ x \in \outstreams(C): \ti{x}{t} \neq \nempty
$$
i.e., $\exists x \in \outstreams(C): \ti{x}{t} \neq \nempty \to \forall y \neq x,\ y \in \outstreams(C):   \ti{y}{t} = \nempty$.
\\
Thus,  a component $C$ is \emph{active} on the time interval $t$ 
if at least one of its output streams is  nonempty on this time interval:
\\
$$
\nactive{C}{t} \ndef \exists x \in \outstreams(C): \nactiveX{C}{t}{x}
$$
and respectively a component $C$ is \emph{restrictively active with a lower/upper bound rb} on the time interval $t$, 
$rb \le \| \outstreams(C) \|$, 
if  on this time interval any $k$  of its output streams are nonempty, where
\begin{itemize}
\item
for lower bound, $\nactiveRL{C}{t}{rb}$: $rb \le k$, i.e. the situation where all of streams are nonempty is allowed, 
and 
\item
for upper bound, $\nactiveRU{C}{t}{rb}$: $k \le rb$, i.e. the situation where all of streams are nonempty is allowed,
\item 
(exact) bound $\nactiveR{C}{t}{rb}$: $k = rb$, i.e. an exact number of streams should be active.
\end{itemize}
$$
\nactiveRL{C}{t}{rb} \ndef   \mid\{x  \in \outstreams(C) \mid \nactiveX{C}{t}{x} \} \mid\ \ge rb  
$$
$$
\nactiveRU{C}{t}{rb} \ndef   \mid\{x  \in \outstreams(C)  \mid \nactiveX{C}{t}{x} \}  \mid\ \le rb  
$$
$$
\nactiveR{C}{t}{rb} \ndef    \mid\{ x \in \outstreams(C)  \mid \nactiveX{C}{t}{x} \}  \mid\  = rb  
$$
If $\forall t: \nactiveR{C}{t}{1}$ we have the case where all the output streams of the component $C$ are \emph{disjoint}.

In a similar way we specify predicates over a set $S$ of components 
to express that on the time interval $t$ some of the components from this set are active: 
$$
\nactiveSet{S}{t} \ndef  \exists C \in S:~ \nactive{C}{t} 
$$
$$
\nactiveSetRL{S}{t}{rb}  \ndef  \mid\{ C \in S  \mid \exists x  \in \outstreams(C): \nactiveX{C}{t}{x} \}  \mid\  \ge rb
$$
$$
\nactiveSetRU{S}{t}{rb}  \ndef    \mid\{ C \in S  \mid \exists x  \in \outstreams(C): \nactiveX{C}{t}{x} \}  \mid\  \le rb
$$
$$
\nactiveSetR{S}{t}{rb} \ndef  \mid\{ C \in S  \mid \exists x  \in \outstreams(C): \nactiveX{C}{t}{x} \}  \mid\  = rb  
$$
$$
\nactiveSetRLC{S}{t}{rb}  \ndef  \mid\{ C \in S  \mid \nactive{C}{t}  \}  \mid\  \ge rb
$$
$$
\nactiveSetRUC{S}{t}{rb}  \ndef    \mid\{ C \in S  \mid \nactive{C}{t}  \}  \mid\  \le rb
$$
$$
\nactiveSetRC{S}{t}{rb} \ndef  \mid\{ C \in S  \mid \nactive{C}{t}  \}  \mid\  = rb  
$$


\subsection{Composition of processes}

\begin{figure}[ht!]
\includegraphics[scale=0.55]{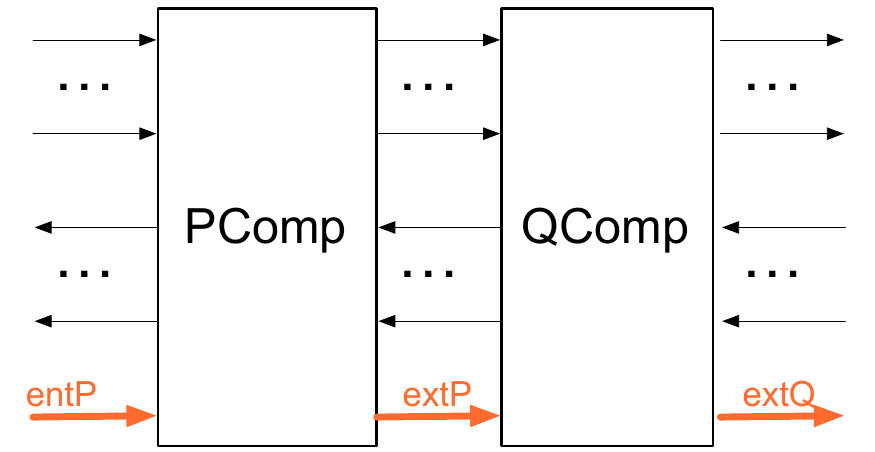}\hspace{2mm}\includegraphics[scale=0.55]{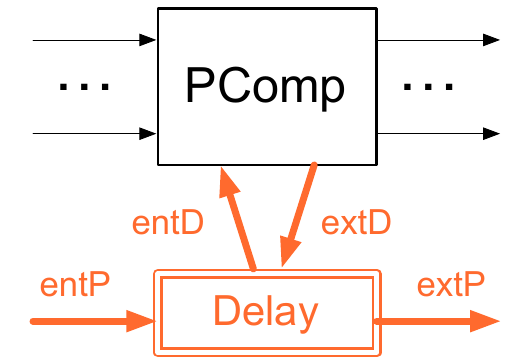}
\\{\footnotesize{(a) Sequential Composition $P$;$Q$}}\hspace{15mm}{\footnotesize (b) Repetitively Comp. Process}
\\~\\~\\
\includegraphics[scale=0.55]{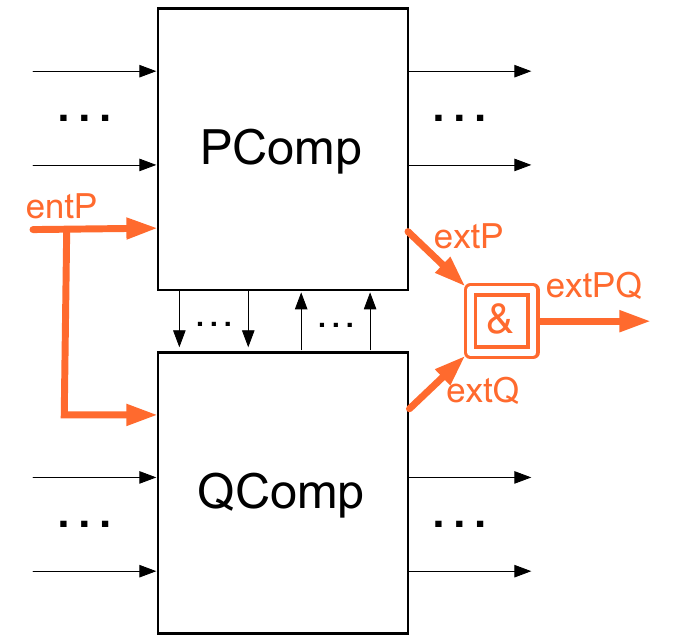}\hspace{1mm}\includegraphics[scale=0.55]{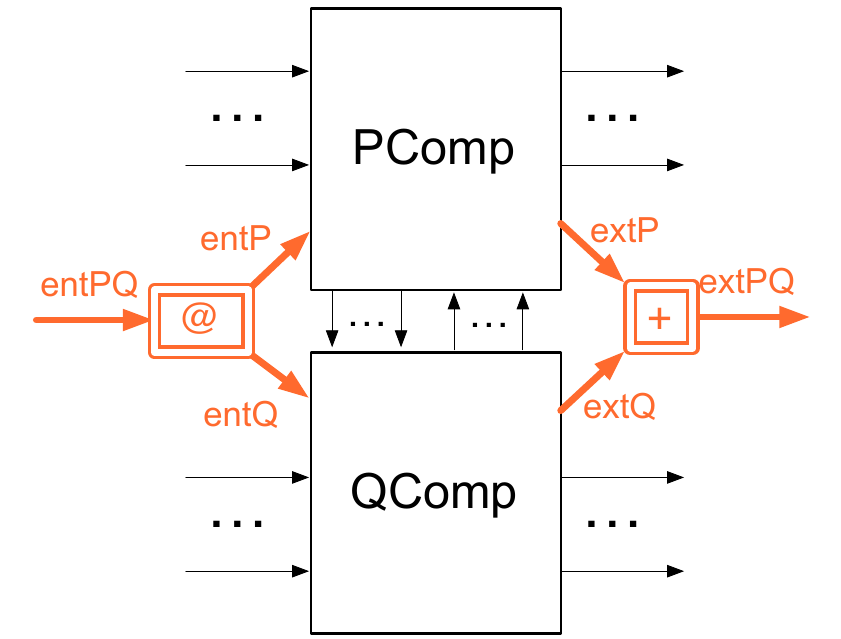}
\\{\footnotesize{(c) Simultaneous Composition $P||Q$}}\hspace{5mm}{\footnotesize (d) Alternative Composition $P\oplus Q$}
\begin{center}
\caption{Composition of Processes $P$ and $Q$}
\label{fig:composition}
\end{center}
\end{figure}

Assume $P$ and $Q$ be any two processes. 
The sets of input and output channels are defined for processes $P$ and $Q$ 
as well as for the the corresponding components $PComp$ and $QComp$, i.e. the component representation of these processes, 
 $PComp = \semantics{P}$ and $QComp = \semantics{Q}$,  as follows:

\[ 
\begin{array}{lcl}
Entry(P) = entP 		&& Entry(Q) = entQ 	
\\
Exit(P) = extP		&& Exit(Q) = extQ			
\\ 
I_P = i_1, \dots, i_m	&& I_Q = x_1, \dots, x_k		
\\%
O_P = o_1,  \dots, o_n	&& O_Q = y_1,  \dots, y_z	
\\%
\end{array}
\] 
\noindent
A general graphical representation of composition is presented on Figure~\ref{fig:composition}. 
All the channels  representing entry and exit points of a process 
(as well as connectors to merge and to split the streams over these channels) are drawn in orange. 
The details of auxiliary component specifications are omitted in this paper, cf. the technical report~\cite{spichkova_processes}.
Having this representation we can analyse properties of composed processes 
by applying a well-developed composition theory, elaborated by Broy \cite{broy_refinements}. 

Among other factors, the purposed representation gives a basis for a straightforward analysis of the worst case execution time (WCET) of the composed processes, e.g., 
it is easy to see that\\
 \emph{wcet}($P$;$Q$) = \emph{wcet}($P$) + \emph{wcet}($Q$), \\ 
 \emph{wcet}($P\circlearrowleft_{lpspec}$) = \emph{wcet}($P$), \\ 
\emph{wcet}($P||Q$) = \emph{max}$\{$\emph{wcet}($P$), \emph{wcet}($Q$)$\}$ + \emph{wcet}($\&$), \\
\emph{wcet}($P \oplus Q$) = \emph{max}$\{$\emph{wcet}($P$), \emph{wcet}($Q$)$\}$ + \emph{wcet}(@) + \emph{wcet}($+$), \\
where \emph{wcet}($X$) denotes the WCET of the process $X$. 
Consequently, on some abstraction level the the WCET of the components $\&$, @ and + can be treated as 0.

 \newpage
 \noindent
\emph{Sequentially composed process $P$;$Q$ (cf. Figure~\ref{fig:composition}a)}
 is the simplest variant of the process composition, which requires no additional auxiliary components:
 \\~
 \\
 \indent
$\semantics{P(i_1, \dots, i_m, o_1,  \dots, o_n) ; Q(x_1, \dots, x_k, y_1,  \dots, y_z)} ~=$\\
\indent
$PComp (entP, i_1, \dots, i_m, extP, o_1,  \dots, o_n) ~\wedge $ \\
\indent
$QComp(extP, x_1, \dots, x_k, extQ, y_1,  \dots, y_z)$ 
\\

\noindent
The entry and exit points are defined in this case by\\
$Entry(P;Q) = entP$ and 
$Exit(P; Q) = extQ$.

~\\ 
\emph{Repetitively composed process (cf. Figure~\ref{fig:composition}b)} 
can be realised in two versions, an autonomous  and a non-autonomous one. 
For both cases, the special component $Delay$ can be defined in many ways to fulfil the required restart-properties, 
however, in most cases it should represent either a timer or a counter of some kind. 
The important point is here that it should be \emph{strict causal}, i.e. to have at least one time unit delay, to prevent Zeno runs~\cite{ZenoRuns} 
 for the case the process $P$   is only weak causal. 
 
 \noindent
In the autonomous version,  the entry and the exit points   are undefined,  
because the process is started by itself and repeated after the time specified by the $Delay$ component:\\
~\\
$
\semantics{P(i_1, \dots, i_m, o_1,  \dots, o_n)\circlearrowleft_{lpspec}^A } ~=$\\
$PComp(entD, i_1, \dots, i_m, extD, o_1,  \dots, o_n) \wedge   Delay(extD, entD)$
\\
~\\
In the non-autonomous version, the $Delay$ component should be specified in more sophisticated way 
to model not only a delay but also react to the start signals from outside, as well as to define whether
  the process can be restarted before it was completed. 
  Thus,  $Entry(P\circlearrowleft_{lpspec}) = entP$ and $Exit(P\circlearrowleft_{lpspec} ) = extP$.\\~
  \\
  $
\semantics{P(i_1, \dots, i_m, o_1,  \dots, o_n)\circlearrowleft_{lpspec}^A } ~=$\\
$PComp(entD, i_1, \dots, i_m, extD, o_1,  \dots, o_n) ~\wedge  \\
 Delay(entP, extD, extP, entD)$

~\\ 
\emph{Simultaneously composed process $P||Q$}  (cf. Figure~\ref{fig:composition}c) 
requires an auxiliary components to join the output control streams, and 
and assumes  that the processes $P$ and $Q$ can be activated next time only in the case when both of them are completed, 
 $Entry(P;Q) = entP$ and $Exit(P; Q) = extPQ$:\\
~
\\
 \indent
$\semantics{P(i_1, \dots, i_m, o_1,  \dots, o_n) || Q(x_1, \dots, x_k, y_1,  \dots, y_z) } ~=$\\
 \indent
 $ PComp(entP, i_1, \dots, i_m, extP, o_1,  \dots, o_n) ~\wedge$\\
 \indent
 $QComp(entP, x_1, \dots, x_k, extQ, y_1,  \dots, y_z) ~\wedge$\\
 \indent
 $ \&(extP, extQ, extPQ)$

\begin{figure}
 \centering
{\scriptsize
\begin{specF}{\&}{}{timed}
\InOut{x, y: Event}
{z: Event}
\scnlocal xReady, yReady: \Bool
\scninit xReady = \fnfalse; yReady = \fnfalse\\
\zeddashline
\uasm \t1
\fmsg{1}{x} ~\wedge~ \fmsg{1}{y}\\   
\ngar \\
\includegraphics[scale=0.6]{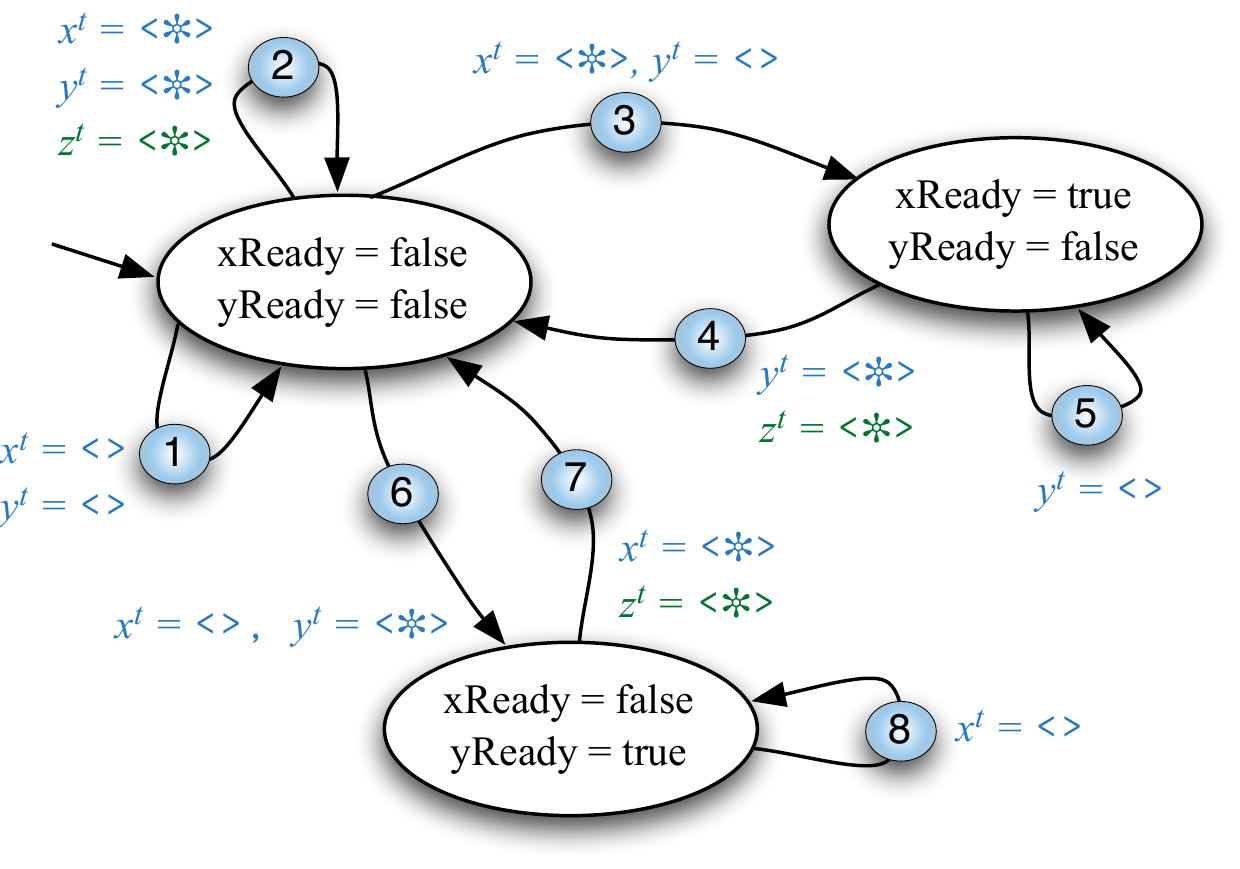}
\end{specF}
}
\vspace{-7mm}
    \caption{Specification of the connector \&}
    \label{fig:process}
\end{figure}

\noindent
The connector $\&$   models the following behaviour: the exit point of $\semantics{P || Q}$  
must be activated iff both processes, $P$ and $Q$ have terminated either simultaneously or one after another. 
Its local variables \emph{xReady} and \emph{yReady}  indicate whether the corresponding  process have already terminated. 
If one of these processes terminates first,  the component 
sets the corresponding  variable  to $\fntrue$   to indicate that the component is waiting for the termination of the second process. 
Only when another process terminates (or if $P$ and $Q$ terminate in the same time unit), the component $\&$ produces the exit-message and set both  variables to $\fnfalse$.

\noindent
\emph{Alternate process $P \oplus Q$} (cf. Fig.~\ref{fig:composition}d, $Exit(P \oplus Q) = extPQ$ and $Entry(P \oplus Q) = entPQ$) 
requires two connectors: @ 
to choose which of the processes should be started (at which process should be sent the activation signal),
 and + to merge the output control flow: \\
 ~\\
 \indent
$\semantics{P(i_1, \dots, i_m, o_1,  \dots, o_n) \oplus Q(x_1, \dots, x_k, y_1,  \dots, y_z)} ~=$\\
 \indent
$P^{spec}(entP, i_1, \dots, i_m, extP, o_1,  \dots, o_n) ~\wedge$\\
 \indent
$Q^{spec}(entQ, x_1, \dots, x_k, extQ, y_1,  \dots, y_z) ~\wedge$\\
 \indent
$\text{@}(entPQ, entP, entQ) ~\wedge~  \text{+}(extP, extQ, extPQ)$
\\
We omit the technical details of the specifications of these connectors in this paper.
%

\section{Conclusions and Future Work}

This paper introduces 
 a formal model of processes that is compatible with the component/ data flow view.   
 This approach reflects   general constrains of the IEC 61499 standard and can can be seen as a formal representation  of  its main ideas. 
 Moreover, it allows to swap from an event-based specification to a  time-triggered one.  
To present our theory of process modelling, we discussed how a process can be represented by a  component as well as 
which properties have the different kinds of composition operators. 

This approach is based on human factor analysis within formal methods~\cite{Spichkova2013HFFM,hffm_spichkova}, 
allows to have short and at the same time readable specifications, 
 and is appropriate for the case the switching to another language is required 
 as well as for application of the specification and proof methodology  
 \cite{spichkova2013we} aligned  on 
the future proofs already during specification phase
to make them simpler and appropriate for application not only in theory but also in practice.

Future research direction comprises  extension of the presented approach by parameterised contracts 
and reliability as well as timing analysis to concurrent systems, 
combining the results introduced in this paper 
with analysis of the WCET  
of a specified process as discussed in~\cite{Lednicki2013}  
as well as 
with prior work in this direction \cite{Fredriksson-etal:2007b}, 
where timing analysis to concurrent systems of both 
WCET in industry-strength tools for large software systems
in distributed control, and of sampled performance in large-scale runs, were analysed.  

~\\
\emph{Acknowledgments:}
We would like to thank  Christian Leuxner, BMW Group, 
               for numerous discussions on the subject of this paper.

\bibliographystyle{IEEEtran}

\begin{thebibliography}{10}
\providecommand{\url}[1]{#1}
\csname url@samestyle\endcsname
\providecommand{\newblock}{\relax}
\providecommand{\bibinfo}[2]{#2}
\providecommand{\BIBentrySTDinterwordspacing}{\spaceskip=0pt\relax}
\providecommand{\BIBentryALTinterwordstretchfactor}{4}
\providecommand{\BIBentryALTinterwordspacing}{\spaceskip=\fontdimen2\font plus
\BIBentryALTinterwordstretchfactor\fontdimen3\font minus
  \fontdimen4\font\relax}
\providecommand{\BIBforeignlanguage}[2]{{%
\expandafter\ifx\csname l@#1\endcsname\relax
\typeout{** WARNING: IEEEtran.bst: No hyphenation pattern has been}%
\typeout{** loaded for the language `#1'. Using the pattern for}%
\typeout{** the default language instead.}%
\else
\language=\csname l@#1\endcsname
\fi
#2}}
\providecommand{\BIBdecl}{\relax}
\BIBdecl

\bibitem{IEC614991}
{IEC 61499-1}, ``{Function blocks - Part 1: Architecture},'' {International
  Electrotechnical Commission}, 2005.

\bibitem{Vyatkin2011_IEC61499}
V.~Vyatkin, ``{IEC 61499 as Enabler of Distributed and Intelligent Automation:
  State-of-the-Art Review},'' \emph{IEEE Trans. Industrial Informatics},
  vol.~7, no.~4, pp. 768--781, 2011.

\bibitem{ReisigPetriNets}
W.~Reisig, \emph{Petri nets: an introduction}.\hskip 1em plus 0.5em minus
  0.4em\relax Springer, 1985.

\bibitem{Salimifard}
K.~Salimifard and M.~Wright, ``Petri net-based modelling of workflow systems:
  An overview,'' \emph{European Journal of Operational Research}, vol. 134,
  no.~3, pp. 664--676, 2001.

\bibitem{CBSQ2003}
A.~Cechich, M.~Piattini, and A.~Vallecillo, Eds., \emph{{Component-Based
  Software Quality: Methods and Techniques}}, ser. LNCS.\hskip 1em plus 0.5em
  minus 0.4em\relax Springer, 2003, vol. 2693.

\bibitem{Broy2010MSS}
M.~Broy, ``Multifunctional software systems: Structured modeling and
  specification of functional requirements,'' \emph{Sci. Comput. Program.},
  vol.~75, no.~12, pp. 1193--1214, 2010.

\bibitem{cocome}
M.~Broy, J.~Fox, F.~H\"{o}lzl, D.~Koss, M.~Kuhrmann, M.~Meisinger,
  B.~Penzenstadler, S.~Rittmann, B.~Sch\"{a}tz, M.~Spichkova, and D.~Wild,
  ``{Service-Oriented Modeling of CoCoME with Focus and AutoFocus},'' pp.
  177--206, 2008.

\bibitem{Kapova2010}
L.~Kapova, B.~Buhnova, A.~Martens, J.~Happe, and R.~Reussner, ``State
  dependence in performance evaluation of component-based software systems,''
  in \emph{{Performance engineering}}.\hskip 1em plus 0.5em minus 0.4em\relax
  ACM, 2010, pp. 37--48.

\bibitem{CB-SPE}
A.~Bertolino and R.~Mirandola, ``Cb-spe tool: Putting component-based
  performance engineering into practice,'' in \emph{Component-Based Software
  Engineering}, ser. LNCS, I.~Crnkovic, J.~Stafford, H.~Schmidt, and
  K.~Wallnau, Eds., vol. 3054.\hskip 1em plus 0.5em minus 0.4em\relax Springer,
  2004, pp. 233--248.

\bibitem{Robocop}
E.~Bondarev, P.~de~With, and M.~Chaudron, ``Predicting real-time properties of
  component-based applications,'' in \emph{In Proc. of the 30the EUROMICRO
  conference}, 2004, pp. 40--47.

\bibitem{Becker06performanceprediction}
S.~Becker, L.~Grunske, R.~Mirandola, and S.~Overhage, ``Performance prediction
  of component-based systems: A survey from an engineering perspective,'' in
  \emph{Architecting Systems with Trustworthy Components}, ser. LNCS, vol.
  3938.\hskip 1em plus 0.5em minus 0.4em\relax Springer, 2006, pp. 169--192.

\bibitem{UML_2005}
I.~J. G.~Booch, J.~Rumbaugh, \emph{Unified Modeling Language User Guide}.\hskip
  1em plus 0.5em minus 0.4em\relax Addison-Wesley Professional, 2005.

\bibitem{Storrle04semanticsand}
H.~St{\"o}rrle, ``Semantics and verification of data flow in uml 2.0
  activities,'' in \emph{ENTCS}.\hskip 1em plus 0.5em minus 0.4em\relax
  Elsevier, 2004, pp. 35--52.

\bibitem{reo2012}
N.~Kokash, C.~Krause, and E.~de~Vink, ``{Reo + mCRL2: A framework for
  model-checking dataflow in service compositions},'' \emph{Formal Aspects Of
  Computing}, vol.~24, no.~2, pp. 187--216, 2012.

\bibitem{Aalst03yawl_yet}
W.~van~der Aalst and A.~H. M.~T. Hofstede, ``{YAWL: Yet Another Workflow
  Language},'' \emph{Information Systems}, vol.~30, pp. 245--275, 2003.

\bibitem{Schmidt2001trustByContracts}
H.~Schmidt, I.~Poernomo, and R.~Reussner, ``Trust-by-contract: Modelling,
  analysing and predicting behaviour of software architectures,'' \emph{J.
  Integr. Des. Process Sci.}, vol.~5, no.~3, pp. 25--51, Aug. 2001.

\bibitem{HandbookProcessAlgebra}
J.~A. Bergstra, \emph{Handbook of Process Algebra}, A.~Ponse and S.~A. Smolka,
  Eds.\hskip 1em plus 0.5em minus 0.4em\relax Elsevier Science Inc., 2001.

\bibitem{ACP}
J.~A. Bergstra and J.~W. Klop, ``Algebra of communicating processes with
  abstraction,'' \emph{Theor. Comput. Sci.}, vol.~37, pp. 77--121, 1985.

\bibitem{CSP}
C.~A.~R. Hoare, ``Communicating sequential processes,'' \emph{Commun. ACM},
  vol.~21, no.~8, pp. 666--677, Aug. 1978.

\bibitem{CCS}
R.~Milner, \emph{A Calculus of Communicating Systems}, ser. LNCS.\hskip 1em
  plus 0.5em minus 0.4em\relax Springer, 1980, vol.~92.

\bibitem{hilderink2003}
G.~H. Hilderink, ``Graphical modelling language for specifying concurrency
  based on {CSP},'' \emph{IEEE Software}, vol. 150, pp. 108--120, 2003.

\bibitem{leux2010}
C.~Leuxner, W.~Sitou, and B.~Spanfelner, ``{A formal model for work flows},''
  in \emph{8th IEEE International Conference on Software Engineering and Formal
  Methods (SEFM)}, 2010, pp. 135--144.

\bibitem{Spichkova2013HFFM}
M.~Spichkova, ``{Design of formal languages and interfaces: ``Formal'' does not
  mean ``unreadable''},'' in \emph{Emerging Research and Trends in
  Interactivity and the Human-Computer Interface}, K.~Blashki and P.~Isaias,
  Eds.\hskip 1em plus 0.5em minus 0.4em\relax IGI Global, 2013, pp. 301--314.

\bibitem{focusST}
M.~Spichkova, J.~Blech, P.~Herrmann, and H.~Schmidt, ``{Modeling Spatial
  Aspects of Safety-Critical Systems with FocusST},'' \emph{11th Workshop on
  Model Driven Engineering, Verification and Validation}, 2014.

\bibitem{focus}
M.~Broy and K.~St{\o}len, \emph{Specification and Development of Interactive
  Systems: Focus on Streams, Interfaces, and Refinement}.\hskip 1em plus 0.5em
  minus 0.4em\relax Springer, 2001.

\bibitem{maraninchi03mode}
F.~Maraninchi and Y.~R{\'e}mond, ``Mode-automata: a new domain-specific
  construct for the development of safe critical systems,'' \emph{Sci. Comput.
  Program.}, vol.~46, no.~3, pp. 219--254, 2003.

\bibitem{spichkova_processes}
M.~Spichkova, ``{Focus on processes},'' {TU M{\"u}nchen}, Tech. Report
  TUM-I1115, 2011.

\bibitem{broy_refinements}
M.~Broy, ``Compositional refinement of interactive systems,'' \emph{J. ACM},
  vol.~44, no.~6, pp. 850--891, 1997.

\bibitem{ZenoRuns}
R.~G\'{o}mez and H.~Bowman, ``Efficient detection of zeno runs in timed
  automata,'' in \emph{{Proceedings of the 5th international conference on
  Formal modeling and analysis of timed systems}}.\hskip 1em plus 0.5em minus
  0.4em\relax Springer, 2007, pp. 195--210.

\bibitem{hffm_spichkova}
M.~Spichkova, ``{Human Factors of Formal Methods},'' in \emph{{IADIS Interfaces
  and Human Computer Interaction (IHCI 2012)}}, 2012.

\bibitem{spichkova2013we}
M.~Spichkova, X.~Zhu, and D.~Mou, ``Do we really need to write documentation
  for a system?'' in \emph{International Conference on Model-Driven Engineering
  and Software Development}, 2013.

\bibitem{Lednicki2013}
L.~Lednicki, J.~Carlson, and K.~Sandstr\"{o}m, ``Model level worst-case
  execution time analysis for iec 61499,'' in \emph{{16th International
  Symposium on Component-based software engineering}}.\hskip 1em plus 0.5em
  minus 0.4em\relax ACM, 2013, pp. 169--178.

\bibitem{Fredriksson-etal:2007b}
J.~Fredriksson, T.~Nolte, M.~Nolin, and H.~Schmidt, ``Contract-based reusable
  worst-case execution time estimate,'' in \emph{Proceedings of the 13th IEEE
  International Conference on Embedded and Real-Time Computing Systems and
  Applications}, 2007, pp. 39--46.

\end{thebibliography}


\end{document}